\documentclass[conference]{IEEEtran}

\usepackage{cite}
\usepackage{color}
\usepackage{multirow}
\usepackage{booktabs,multirow,array}
\usepackage[table]{xcolor} 
\usepackage{ltxcmds}
\usepackage{footnote}
\usepackage{amsthm}

\usepackage[table]{xcolor}
\usepackage{subfigure}
\usepackage{lipsum}
\usepackage{amssymb} 
\usepackage{pdflscape}
\usepackage{afterpage}

\usepackage{balance}

\usepackage{pifont}
\newcommand{\xmark}{\ding{55}}%

\ifCLASSINFOpdf
\usepackage[pdftex]{graphicx}
\else
  \usepackage[dvips]{graphicx}
\fi

\begin{document}
\title{Addressing Adversarial Machine Learning Attacks in Smart Healthcare Perspectives }

\author{\IEEEauthorblockN{Arawinkumaar Selvakkumar$^{\ast}$, Shantanu Pal$^{\ast}$, Zahra Jadidi$^{\ast\mathsection}$}
\IEEEauthorblockA{$^{\ast} $School of Computer Science, Queensland University of Technology, Brisbane, QLD 4000, Australia\\ $^{\mathsection} $School of Information and Communication Technology, Griffith University, Gold Coast Campus, QLD 4222, Australia\\
{arawinkumaar.selvakkumar@connect.qut.edu.au,
shantanu.pal@qut.edu.au,
zahra.jadidi@qut.edu.au} {}}}




\maketitle


\begin{abstract}

Smart healthcare systems are gaining popularity with the rapid development of intelligent sensors, the Internet of Things (IoT) applications and services, and wireless communications. However, at the same time, several vulnerabilities and adversarial attacks make it challenging for a safe and secure smart healthcare system from a security point of view. Machine learning has been used widely to develop suitable models to predict and mitigate attacks. Still, the attacks could trick the machine learning models and misclassify outputs generated by the model. As a result, it leads to incorrect decisions, for example, false disease detection and wrong treatment plans for patients. In this paper, we address the type of adversarial attacks and their impact on smart healthcare systems. We propose a model to examine how adversarial attacks impact machine learning classifiers. 
To test the model, we use a medical image dataset. Our model can classify medical images with high accuracy. We then attacked the model with a Fast Gradient Method Sign attack (FGSM) to cause the model to predict the images and misclassify them inaccurately. Using transfer learning, we train a VGG-19 model with the medical dataset and later implement the FGSM to the Convolutional Neural Network (CNN) to examine the significant impact it causes on the performance and accuracy of the machine learning model. Our results demonstrate that the adversarial attack misclassifies the images, causing the model's accuracy rate to drop from 88\% to 11\%. 

\end{abstract}

\IEEEpeerreviewmaketitle

\section{Introduction}
\label{introduction}

Machine learning has helped multiple industries like healthcare meet their growing demands and exceed expectations in automation~\cite{ahmad2018interpretable}. It helps streamline administrative processes and allows building models to quickly and efficiently deliver results to analyse the patient's critical data~\cite{qayyum2020secure}. According to~\cite{stasha2021}, 92\% of different healthcare solutions promote digitisation for healthcare facilities more efficiently in their services. However, current challenges include using emerging technologies, e.g., machine learning, and better utilising their health data processing~\cite{mustafa2021automated}.  

With the growth of intelligent sensors, the Internet of Things (IoT) applications and services \cite{rabehaja2019design}, advanced wireless communication technologies, and Artificial Intelligence (AI) introduced new security challenges for the smart healthcare systems impacting the performance of a machine learning model \cite{xiao2018iot}. For example, a new attack dimension, known as adversarial attacks could impact the machine learning classifiers used in a smart healthcare setting~\cite{churcher2021experimental}. 
Adversarial attacks are significant as the inaccurate predictions could lead to incorrect disease detection and wrong treatment plans for patients, which can cause false diagnoses~\cite{castiglioni2021ai}.

Adversarial attacks, also known as adversarial machine learning, are cyber-attacks that attempts to trick a machine learning model into producing incorrect predictions and causing the model to malfunction~\cite{huang2011adversarial}. Adversarial machine learning attacks are originated when a perturbed input is injected into an image. 
Deep Neural Networks (DNNs) have showcased their superior performance in visual understanding tasks like image classification and video recognition~\cite{jiang2019black}. 
Adversarial attacks could impact both image and video elements of data, causing the machine learning model to make an inaccurate prediction with a high level of confidence~\cite{akhtar2021advances}. Commonly, a majority of the attacks in the healthcare domain target medical imaging data (e.g., X-rays, CT scans, radiographs, MRI's, etc.) to alter the predicted disease~\cite{newaz2020adversarial}. 
Therefore, we intend to focus on misclassifying images using adversarial attacks. 

Several proposals discuss the impact of security and attacks in smart healthcare systems \cite{newaz2021survey} \cite{islam2015internet} \cite{pal2018policy} \cite{pal2018fine}. A few of them also present vulnerabilities in machine learning models and the type of adversarial attacks that could impact smart healthcare systems~\cite{li2021comprehensive}~\cite{pitropakis2019taxonomy}. While these proposals try to address the types of adversarial attacks and vulnerabilities of machine learning models, they do not explicitly discuss the significant impact of adversarial attacks on machine learning models, particularly in machine learning classifiers. Moreover, those proposals do not address the implementation of adversarial attacks to a Convolutional Neural Network (CNN)~\cite{li2021survey} that have been trained with medical images.

\begin{table*}[ht]
\small
\centering
\caption{Previous proposals on adversarial attacks on machine learning and their comparison with our work.}
\label{tab:comparison-previous-works}

\scalebox{1}{
\begin{tabular}{c*{5}{c}r}
\hline  

Reference & Smart Healthcare & Addressed Adversarial & Implementation of & Implementation of & Use of \\ {} & Context & Attacks & CNN & FGSM & Custom Dataset \\[0.5ex] 
\hline

\cite{newaz2020adversarial} & \checkmark & \checkmark & \xmark & \checkmark & \checkmark \\ [0.5ex]

\cite{anthi2021adversarial} & \xmark & \checkmark & \xmark & \checkmark & \checkmark \\ [0.5ex] 

\cite{rosenberg2020adversarial} & \xmark & \checkmark & \xmark & \checkmark & \checkmark \\ [0.5ex] 

\cite{ren2020adversarial} & \xmark & \checkmark & \xmark & \checkmark & \xmark  \\ [0.5ex] 

\cite{ibitoye2019threat} & \xmark & \checkmark & \xmark & \xmark & \xmark  \\ [0.5ex]

\cite{goodfellow2014explaining} & \xmark & \checkmark & \xmark & \checkmark & \xmark  \\ [0.5ex]

\cite{tian2019smart} & \checkmark & \xmark & \xmark & \xmark & \xmark  \\ [0.5ex]

\cite{finlayson2019adversarial} & \checkmark & \checkmark & \xmark & \xmark & \xmark  \\ [0.5ex]

\cite{chen2020hopskipjumpattack} & \xmark & \checkmark & \xmark & \checkmark & \xmark  \\ [0.5ex]

\textbf{[Our Work]} & \checkmark & \checkmark 
& \checkmark & \checkmark & \checkmark \\ [0.5ex] 

\hline
\end{tabular}
}
\end{table*}

In this paper, we motivate to examine the impact of adversarial machine learning attacks against smart healthcare systems, particularly for the machine learning classifiers. We propose a method that focuses on the Fast Gradient Sign Method (FGSM)~\cite{liu2019sensitivity} attack and how an adversarial attack can be implemented to a CNN. Furthermore, our proposed method tries to mislead the machine learning classifiers by implementing the FGSM attack to the CNN and examining the overall impact it caused on the network's performance. 
The major contributions of the paper are:

\begin{itemize}
    \item We propose a method to detect adversarial attacks to identify the flaws that could impact smart healthcare systems. Our method considers the attack in the training and testing phase and shows how they can potentially mislead the system.
    
    \item  We use a real-world medical image dataset to test our system. Using transfer learning, we trained a VGG-19 model \cite{wen2019new} with the dataset to correctly classify the images. 
    
    \item We implement one white-box attack, Fast Gradient Sign Method (FGSM) attack, to a Convolutional Neural Network (CNN) to investigate the impact it causes to the model.
    
    \item We evaluate the proposed attack to the system by using a model classification report and a confusion matrix of the testing data.
\end{itemize}

The rest of the paper is organised as follows. In Section~\ref{related-work}, we discuss related work. In Section~\ref{method}, we present the methods and models selected for experimental analysis. Section~\ref{results} discuss the results. Finally, Section~\ref{conclusion} concludes the paper with future work.

\section{Related Work}
\label{related-work}

Several studies discuss adversarial attacks and their impact on machine learning models. In Table~\ref{tab:comparison-previous-works}, we summarise the existing works and show their comparison with our work. For instance, proposal~\cite{newaz2020adversarial} provides an analysis of adversarial attacks by utilising five adversarial algorithms that are used to examine certain attack models.  However, the dataset used in~\cite{newaz2020adversarial} is generated from device data. Moreover, unlike our proposal, they do not use a CNN to test the model's accuracy. 

The authors in~\cite{anthi2021adversarial} explore the consequences of adversarial attacks and show how adversarial training can support the
robustness of supervised models. They use confusion matrices and
classification performance of J48 and Random Forest models
trained on adversarial samples generated. Notably, \cite{anthi2021adversarial} addresses the unique challenges of implementing adversarial attacks in the cybersecurity domain. Although the paper discusses the FGSM attack, unlike our proposal, it does not use CNN. 

Proposal~\cite{rosenberg2020adversarial} presents an overview of different adversarial attacks, including FGSM. However, unlike our motivation, they use the MNIST and CIFAR10 datasets, not a custom dataset. Proposal~\cite{ren2020adversarial} discusses theoretical foundations, algorithms, and applications of adversarial attack techniques along with defence techniques to combat them and protect systems. Unlike our proposal, it does not consider an implementation detail. Proposals~\cite{ibitoye2019threat} and~\cite{goodfellow2014explaining} discuss the issues of machine learning models misclassifying, the FGSM attack, and adversarial training. But once gain, unlike our approach, they do not consider CNN. 

In~\cite{tian2019smart}, authors discuss adversarial machine learning from a smart healthcare perspective. However, unlike ours, the proposal does not consider FGSM and CNN, nor any detailed design is stated. While proposals~\cite{finlayson2019adversarial} and~\cite{chen2020hopskipjumpattack} present some technical details on implementing adversarial attacks on machine learning, unlike ours, they do not use a custom dataset. Further, they do not focus on the smart healthcare perspective. 

In summary, we note that the above-mentioned proposals do
not discuss the implementation of a CNN. While a few proposals discuss the importance of CNN in adversarial machine learning attacks, the discussion does not lead to an actual execution that we provide in our paper. That said, we focus on implementing an FGSM attack on CNN. We also note there is a trend to use an existing dataset, e.g., MNIST and CIFAR-10. In our paper, we avoid such datasets to show the actual implementation of the model using a medical health dataset.

\section{Model and Methods}
\label{method}



In this section, we discuss the model and methods used for the experiment. For more clarity, we separate the experiment into two parts. In Part 1 (cf. Section~\ref{part-1}), we train a VGG-19 model with a medical dataset of~\cite{dataset}, containing images of both benign and malignant skin cancer moles, 
 and, in Part 2 (cf. Section~\ref{part-2}), the model has been trained, and the FGSM attack is generated and implemented to the trained model to examine the impact.

\subsection{Experimental Setup}

All training, testing and evaluation were performed on a laptop running Microsoft Windows 10 Home. The PC is equipped with an Intel(R) Core (TM) i7-1065G7 processor operating at 1.30GHz and 16GB of RAM. The development environment was Google Colab~\cite{colab}, 
and the two open-source libraries mainly used were Tensorflow (version 2.3.0) and Keras (version 2.4.3). Keras is a deep-learning API written in Python, and Tensorflow is a Python library that can be used to create deep learning models. Python is also used to build CNN and generate adversarial attacks. 
We downloaded the dataset from Kaggle \cite{dataset}. A detailed description of the employed dataset is given below.

\subsection{Part 1 – Training the CNN with a Medical Dataset}
\label{part-1}

\subsubsection{Dataset}

It contains skin cancer images of the International Skin Imaging Collaboration (ISIC) 
Archive~\cite{dataset}. The dataset was last updated on 20-06-2019. It has a balanced number of images of benign and malignant skin moles. The dataset has a total of 3,297 images with the \textit{.jpg} file extension and the 224$\times$224 image dimension. The dataset has two folders, benign and malignant, which separates the images into two categories. For our experiment, 500 images were selected at random 
from the benign and malignant folders and downloaded. The skin cancer dataset was collected to simulate a disease detection application in a smart healthcare system. This dataset is useful because we can use the images to build a disease detection application, where the model can predict whether the images of skin cancer moles are benign or malignant. 
The machine learning model 
will classify if the moles are benign or malignant, and it would speed up the efficiency of detecting diseases compared to screening tests which are time-consuming and may require more resources to produce results. 

\subsubsection{Preparing the Dataset}
After the dataset was downloaded from Kaggle, we stored it on Google drive and mounted the dataset in the drive to the Google Colab notebook. Once the necessary libraries and packages were imported, we created an array containing the image label and
data. The images are resized to 224$\times$224 resolution and the number for RGB is 3. Once the array was created, the files were
converted to one-hot encoding. One-hot encoding is a process where categorical variables are converted into integers so the machine learning algorithms can make better predictions. This can be done by importing the LabelEncoder from sklearn’s library. The data is then split into Training and Testing portions using stratified sampling. The training images are split to 70\% while testing images are split to 30\%. To increase the number of relevant images, we use data augmentation to change the present images to expand the dataset. In this case, we set the rotation range of the image to 15. Images that have been rotated, scaled, cropped, and flipped has led to many models misclassifying the images. By using augmented data, it can train the CNN to correctly classify the images although they have been rotated.

\subsubsection{Transfer Learning using VGG-19}

Recall, a CNN is a deep-learning architecture that is used to analyse visual imagery like hand-written digits from the MNIST dataset and images from the CIFAR-10 and custom datasets. 
A CNN contains a combination of layers that transforms an image to an output understood by the model. The layers are the convolutional layer, pooling layer, and three fully connected layers. Note, building a CNN from the ground up takes a long time to train the model and is highly time-consuming~\cite{gu2018recent}. 
We follow a transfer learning approach and use a VGG-19 model, which is an advanced CNN with pre-trained layers. Transfer learning is the learning approach of taking features used to solve one problem and leveraging them on another problem \cite{chollet2020transfer}. 
Moreover, transfer learning can decrease the training time for a neural network model, which can fasten the process. 

The VGG-19 model is a widely used image classification architecture for many datasets. 
It utilises 19 images and has been trained
on millions of image samples. The CNN architecture 
comprises of three layers, the Convolution layer, Max pooling layer and the fully-connected layer. The `Softmax' function is an activation function that is used to
predict the probabilities of each class~\cite{tiwaritransfer}. 
For our experiment, the VGG-19 base model did not include the 3 fully-connected layers at the top because we want to train the model with the pre-trained image weights. 
We have used pre-trained image weights from the ImageNet model. For input shape, it was set to 224,224,3. Then, we froze the base model and flatten the output, which will convert the data into a one-dimensional array for inputting it into the next layer. A common approach to transfer learning would be to add some new trainable layers on
top of the frozen layers, which will turn the old features into predictions on a new dataset. After that, we train the layers on
the new dataset ~\cite{chollet2020transfer}.

For the neural transfer part, we added the pre-trained VGG-19
model and added a dropout of 0.25. Dropout is a regularization
technique for neural network models to prevent the model from
overfitting~\cite{brownlee2016dropout}. 
Once the earlier layers were
frozen, we added the dense layers, which were 128 with `relu'
activation with a dropout of 0.25, 64 with `relu' activation and
a 2 with a `softmax' activation. The model summary shows that there was a total of 23 million parameters with 3 million
trainable and 20 million non-trainable parameters. Trainable
parameters are parameters that were not updated or optimised
during training whereas non-trainable parameters are
parameters that were updated and optimised during training.

\subsubsection{Train the Model}
Once the model is created, we train the model. The learning rate of the model was set to 0.0001, and the EPOCHS,
also known as a number of iterations, was set to 15. The batch size
that selects the number of images that are trained together was
32. The optimiser instantiated was `Adam'. The Adam algorithm can handle sparse gradients on noisy problems and is
easy to configure. The categorical cross-entropy command was
added to the CNN 
as a loss function for multi-class classification when there
are two or more label classes~\cite{keraslosses}. The metric
measured was the model's accuracy so we can determine if the
model is ready to be attacked later. The metric and accuracy, is used
to display the impact on the performance of a machine learning
model after it has been attacked by an adversarial attack. 
The results of the model are discussed in Section~\ref{results}.

\subsection{Part 2 – Implementing FGSM Attack to the Trained CNN}
\label{part-2}

In Part 1, we focused on training the VGG-19 model with a medical dataset to accurately classify the benign and malignant skin-cancer images. In Part 2, we will focus on generating the adversarial attack and implementing it to the system. 


\subsubsection{Generating Image Adversary}
\begin{figure}[ht]
 \centering
    \includegraphics[scale=1.1]{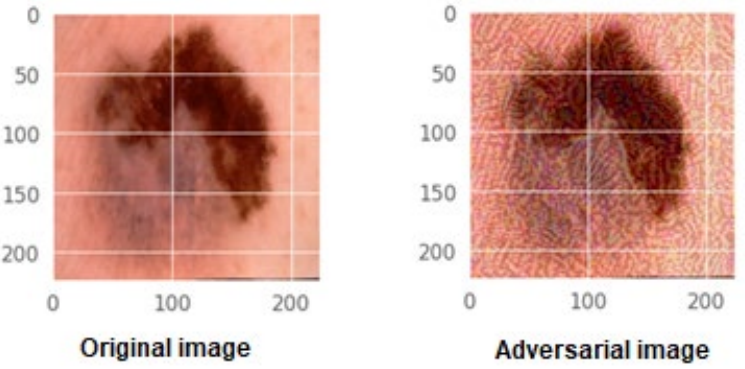}
    \caption{Comparison of the original image and an adversarial image.}
    \label{fig:4}
    \small
\end{figure}

To implement the attack, we first generated an image adversary and recorded our gradients. We used our model to make predictions on the input image and then compute the loss. From
there, we calculate the gradient of loss and compute the sign of the gradient. Then, we construct the image adversary by adding the image with the sign of the gradient. Fig.~\ref{fig:4} illustrates the difference between the original image and an adversarial image from a sample image of the dataset.

After the adversarial image was created, it was sent to the trained model to classify the images. The evaluation of the model after the attack, is reported in Section~\ref{results}. 


\section{Results and Discussion}
\label{results}

In this section, first, we discuss the results achieved from both Part 1 and Part 2 of the methods discussed above. 
Then, we present the performance comparison and the defensive strategies to protect and strengthen the CNN 
from future attacks.  


\subsection{Results: Part 1 of Methods}
After the training was completed, a classification report along
with the training loss and accuracy graph was produced to further understand the changes to the model. In Table~\ref{tab:part1}, we illustrate the
classification results of the trained model. 


\begin{table}[t]
\centering
	\caption{Model's classification results in Part 1}
	\label{table2_pseudo}
	\begin{tabular}{ c c c c c } 
 \hline
  & precision (\%) & recall (\%) & f1-score (\%) & support\\ [0.5ex]
  \hline
  \hline
 Benign & 87 & 90 & 89 & 150 \\[0.5ex]
 Malignant & 90 & 87 & 88 & 150 \\ [0.5ex]
 \hline
\end{tabular}
	\label{tab:part1}%
\end{table}

The classification report visualiser displays (cf. Table~\ref{tab:part1}) the \textit{precision}, \textit{recall}, \textit{f1-score}, and \textit{support} scores of the trained model. The
numbers shown are multiplied by 100 to get percentage values. Precision calculates the model's accuracy of positive predictions. The model has correctly predicted the benign images by 87\% and malignant images by 90\%. Recall is the ability of a classifier to find all positive instances, where benign
was 90\% and malignant was 87\%. 
The f1-score calculates the percentage of positive predictions that were correct, where benign and
malignant were given 89\% and 88\% respectively. Support is the
number of actual occurrences of a class in a specified dataset and
the number is 150 for both benign and malignant labels~\cite{kohli2019understanding}. 

\begin{figure}[ht]
 \centering
    \includegraphics[scale=1.3]{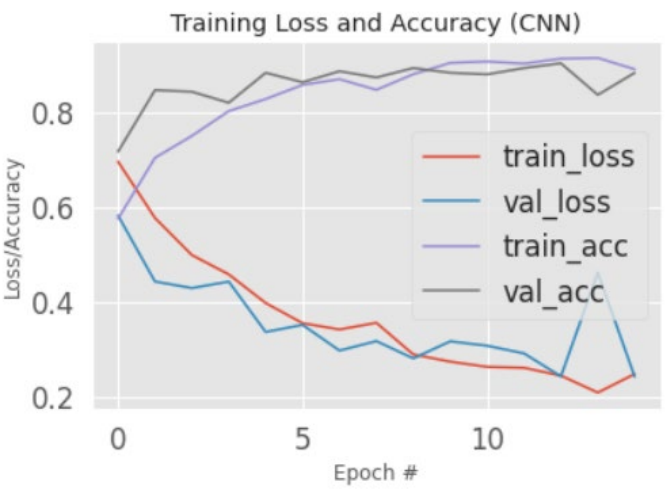}
    \caption{Training Loss and Accuracy graph of the model during training.}
    \label{fig:2}
    \small
\end{figure}


\begin{figure}[ht]
 \centering
    \includegraphics[scale=1]{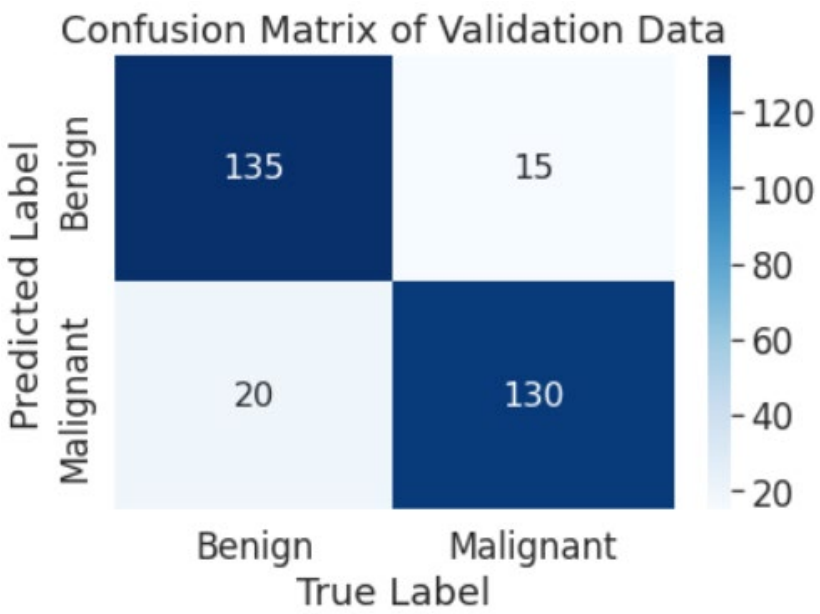}
    \caption{Confusion matrix of the model in Part 1 of Methods.}
    \label{fig:3}
    \small
\end{figure}

A training loss and accuracy graph can visually depict the loss and model accuracy during training.
Fig.~\ref{fig:2} shows the training loss and accuracy graph. The x-axis is the number of epochs/iterations, and the y-axis is the percentage of loss/accuracy. The graph shows that the training loss and validation loss continue to drop as the number of epochs/iterations increase. Moreover, the training accuracy and validation accuracy steadily rise as the number of epochs/iterations increases. The higher the number of epochs, the longer the training time for the CNN. The higher the number of epochs may increase the accuracy of the model. However, the risk of the model overfitting is present.

A confusion matrix was used to evaluate the model's classification accuracy. Fig.~\ref{fig:3} shows the confusion matrix of testing data for the trained model.

The confusion matrix displays the number of images that have been correctly and incorrectly classified by the model. To calculate the \textit{true positive} rate and \textit{true negative} rate, we divided the number of images detected with the total number of images. The true positive rate for `Benign' images was 90\% (135 images) with a false positive rate of 10\% (15 images). In the `Malignant' section, the false negative rate was 13\% (20 images) with a true negative rate of 87\% (130 images). Overall, we can envision that the model has a high classification and
model accuracy.



\begin{figure}[ht]
 \centering
    \includegraphics[scale=.8]{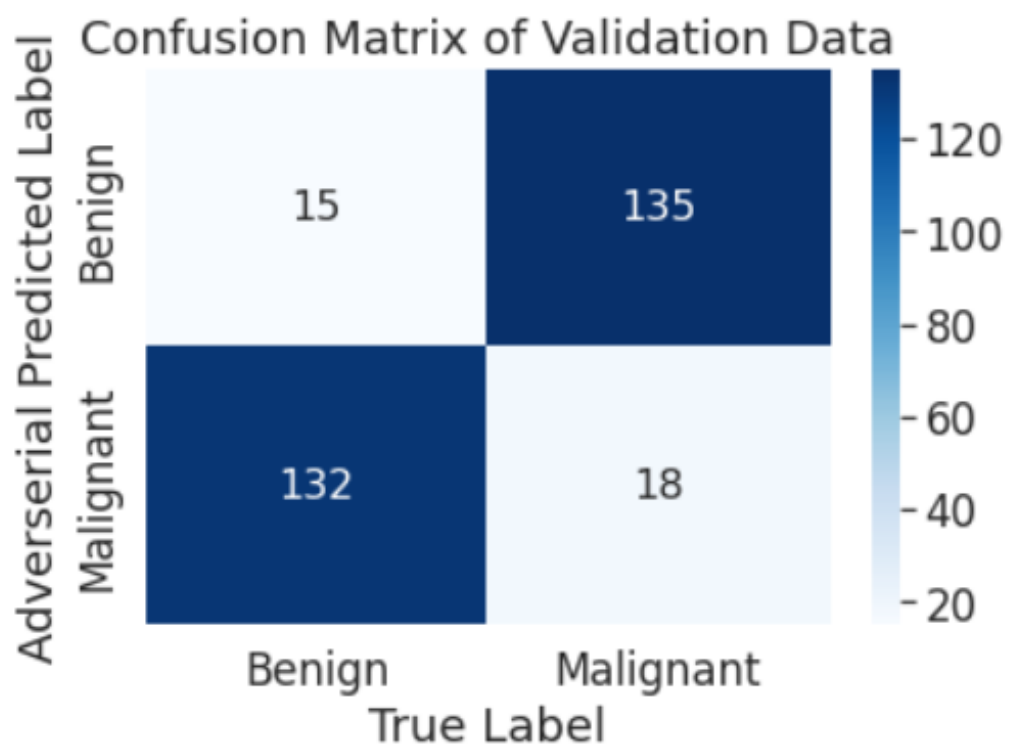}
    \caption{Confusion matrix of the model in Part 2 of Methods.}
    \label{fig:7}
    \small
\end{figure}

\subsection{Results: Part 2 of Methods}
After  the  FGSM attack was implemented, a confusion matrix was produced to further understand the impact caused to the model. Fig.~\ref{fig:7} shows the confusion matrix of the trained model in Part 2 of Methods.

The confusion matrix in Fig.~\ref{fig:7} shows the impact caused by the FGSM adversarial attack in Part 2 of the Methods. The model has severely misclassified the images The true positive rate for
`Benign' images was 10\% (15 images) with a false positive rate
of 90\% (135 images). In the `Malignant' section, the false
negative rate was 88\% (132 images) with a true negative rate of
12\% (18 images). When compared to Part 1, we note that the FGSM attack has significantly impacted the
performance and accuracy of the trained model.



\subsection{Model Accuracy in Part 1 and Part 2 of Methods} 



Accuracy is equal to the sum of true positives and true negatives divided by the total number of images used for training~\cite{ragan2018taking}. From this, we calculated that the model in Part 1 had an accuracy rate of 88\%. Unfortunately, the model in Part 2 that suffered the attack had an accuracy rate of 11\%.

\begin{figure}[ht]
 \centering
    \includegraphics[scale=1.15]{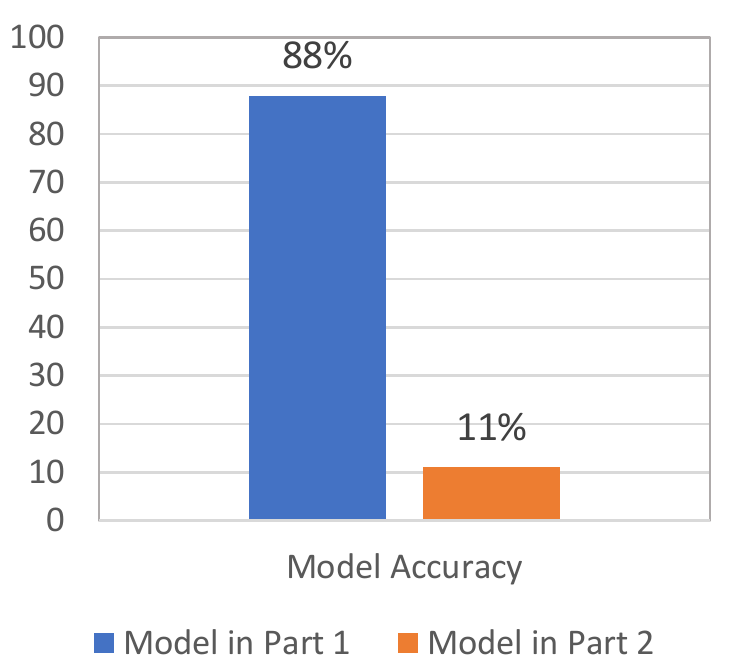}
    \caption{Comparison of model's accuracy of Part 1 and Part 2 of Methods.}
    \label{fig:6}
    \small
\end{figure}

From these findings, we can assess those adversarial attacks, e.g., the FGSM attack used above can significantly downgrade the performance and accuracy of a CNN. The confusion matrix 
has proven that machine learning models are vulnerable to adversarial attacks and has successfully demonstrated our research objectives. Fig.~\ref{fig:6} shows the level of accuracy of the model in Part 1 and Part 2.

In a real-world scenario, the smart healthcare system could be rendered unusable if an adversarial attack tricks the model to
misclassify output, this could
lead to incorrect disease detection and wrong treatment plans for patients in a hospital. The severity of adversarial attacks is high
and could corrupt machine learning models and provide inaccurate information which will affect the services at a hospital. The findings collected are useful as they show the impact an adversarial attack can cause to an image classification model. 


\subsection{Performance Comparison}
In this section, we compare our findings with similar research proposals. For instance, 
The method in~\cite{anthi2021adversarial} comprises of attacking supervised models with adversarial samples. The findings generated was the
classification performance on the adversarial samples generated
and a confusion matrix for the original test set. Although the type
of findings are similar to ours, the results and objectives are different as they used a random forest and J48 classification performance on adversarial samples using the Jacobian-based Sailency Map attack. However, unlike their work, our study focused on using the FGSM attack and creating a confusion matrix that displays the number of images that are predicted correctly or incorrectly. 
Further, proposal~\cite{anthi2021adversarial} evaluated the f1-score on adversarial samples whereas, we evaluated the performance of the model and the classification accuracy. 

The findings in~\cite{paschali2018generalizability} focused towards assessing the
performance of a model using black-box attacks. The authors
reported a comparative evaluation of model robustness and
recorded the average accuracy and classification of the model for both classification and segmentation tasks.
However, their findings form a conclusion that
generalizability and robustness can increase the model's resistance to adversarial examples. 

In addition, proposal~\cite{paschali2018generalizability} implemented three black-box attacks whereas we focused on implementing the FGSM attack to affect the machine learning classifiers, which is missing in~\cite{paschali2018generalizability}. 
Compared to~\cite{newaz2020adversarial}, that evaluated supervised models on adversarial samples, our study aims to examine the impact of adversarial attacks to a deep neural network by evaluating the model's training accuracy and the classification accuracy. 


 \subsection{Defensive Strategies}
\label{defensive-approaches}
 Based on the experimental results, it is
 essential to protect the machine learning model from adversarial attacks to avoid misclassification of images. 
 The two main defensive strategies to combat adversarial attacks in
 the testing stage are \textit{adversarial training} and \textit{randomisation}.
 Adversarial training is a popular defensive approach that
 focuses on training the model with adversarial sample to further
 enhance the model’s accuracy and performance. An evaluation
 by~\cite{goodfellow2014explaining} showed that the
 model can become resistant to adversarial examples once it has
 been trained with adversarial input and produce low error rates.
 Randomisation is the defense strategy to randomise the
 adversarial effects into random effects to mitigate the effects of
 adversarial perturbations in the input. By randomising the
 effects of an adversarial input, it reduces the impact it causes to
 the model. Implementing a defensive strategy using adversarial training is the future aim of this research.

\section{Conclusion and Future Work}
\label{conclusion}
The objective of this paper was to investigate the impact of adversarial attacks on smart healthcare systems. Our findings reported have advanced the initial level of knowledge and understanding by demonstrating how machine learning models are evaluated and the significant impact adversarial attacks can cause to those models.
The findings have shown that machine learning models and CNN are vulnerable to adversarial attacks, e.g., the FGSM attack, and can mislead the model to misclassify outputs. 
The evaluation results of the model’s
accuracy pre-attack and post-attack demonstrated how adversarial attacks could trick machine learning models into misclassifying results. The accuracy of the model before the attack registered at 88\% and it fell to 11\% after being tricked by the adversarial attack. 

The experiment built in this research can be expanded to different areas, e.g., the manufacturing industry for detecting components or faulty appliances, the retail industry for detecting the price and quality of a product and social media platforms where face recognition could be used to
spot users in photos or identify inappropriate content. Although this research is directed to images, videos can be used to build video recognition applications. 

Note, our work has some limitations. For instance, the dataset used to train the machine learning model is relatively smaller, and the number of images could be increased to better train the model. This research is also limited to implementing one adversarial attack instead of more. It would be possible to view different attacks' model accuracy with more adversarial attacks. However, in this paper, we demonstrated a proof of concept and a large-scale deployment could be easily performed as future research work. With further improvement, incorporation of existing deep learning techniques, and defensive approaches, e.g., adversarial training, 
the model has the potential to detect adversarial attacks and prevent them from impacting the model. However, this is a separate direction of research, and we leave this for another future work.

\section*{Acknowledgment}
The authors would like to acknowledge the feedback of Kien Thanh, Queensland University of Technology, Australia. The authors also acknowledge the support of the Commonwealth of Australia and Cybersecurity Research Centre Limited.

\ifCLASSOPTIONcaptionsoff
  \newpage
\fi

\balance{}

\bibliographystyle{IEEEtran}
\bibliography{bare_jrnl}

\end{document}